\documentclass[11pt]{article}

\usepackage{amssymb, latexsym, amsmath}
\usepackage{amsfonts, graphics, amstext}
\usepackage{color}

\def\R{\mathbb R}

\def\N{\mathbb N}

\def\be{\begin{equation}}
\def\ee{\end{equation}}
\def\bea{\begin{eqnarray}}
\def\eea{\end{eqnarray}}
\def\beas{\begin{eqnarray*}}
\def\eeas{\end{eqnarray*}}

\newcommand{\prfe}{\hspace*{\fill} $\Box$
\smallskip \noindent}

\begin{document}

\sloppy

\newtheorem{theorem}{Theorem}[section]
\newtheorem{definition}[theorem]{Definition}
\newtheorem{proposition}[theorem]{Proposition}
\newtheorem{example}[theorem]{Example}
\newtheorem{remark}[theorem]{Remark}
\newtheorem{cor}[theorem]{Corollary}
\newtheorem{lemma}[theorem]{Lemma}

\renewcommand{\theequation}{\arabic{section}.\arabic{equation}}

\title{Rotating, stationary, axially symmetric
        spacetimes with collisionless matter}

\author{H{\aa}kan Andr\'{e}asson\\
        Mathematical Sciences\\
        Chalmers University of Technology\\
        G\"{o}teborg University\\
        S-41296 G\"oteborg, Sweden\\
        email: hand@chalmers.se\\
        \ \\
        Markus Kunze\\
        Mathematisches Institut\\
        Universit\"at K\"oln\\
        Immermannstr.~49-51\\
        D-50931 K\"oln, Germany\\
        email: mkunze@mi.uni-koeln.de\\
        \ \\
        Gerhard Rein\\
        Fakult\"at f\"ur Mathematik, Physik und Informatik\\
        Universit\"at Bayreuth\\
        D-95440 Bayreuth, Germany\\
        email: gerhard.rein@uni-bayreuth.de}

\maketitle

\begin{abstract}
The existence of stationary solutions
to the Einstein-Vlasov system which are
axially symmetric and have non-zero total
angular momentum is shown.
This provides mathematical models for rotating,
general relativistic and asymptotically flat
non-vacuum spacetimes. If angular momentum is allowed to
be non-zero, the system of equations to solve contains
one semilinear elliptic equation which is singular on the
axis of rotation. This can be handled very efficiently
by recasting the equation as one for an axisymmetric
unknown on $\R^5$.
\end{abstract}


\section{Introduction}

\setcounter{equation}{0}

The geometric features of general relativistic and asymptotically
flat spacetimes strongly depend on whether the spacetime has
non-trivial total angular momentum or not.
A well known point in case is the Kerr family, a two parameter family
of stationary vacuum spacetimes which contain a black hole.
The two parameters are the ADM mass ${\cal M}\geq 0$ and the
total angular momentum  ${\cal L}\geq 0$.
If ${\cal L}=0$ one obtains the Schwarzschild spacetime, i.e.,
a static, spherically symmetric black hole of mass ${\cal M}$.
Compared to this the case ${\cal L}>0$ exhibits
a vastly more complicated geometry, and we refer to \cite{Wald} for details.
However, most astrophysical objects are not exactly spherically symmetric,
and many rotate about some axis and have non-trivial total angular momentum.
Hence the mathematical difficulties entailed by giving up spherical symmetry
and allowing for non-zero angular momentum have to be overcome in order
to get closer to physically meaningful models.
There are at the moment only two papers where
the existence of rotating equilibrium configurations of self-gravitating matter
distributions is shown in the framework of General Relativity:
these are \cite{HeiligGR} and \cite{ABS09},
where matter is modeled as an ideal fluid and as an elastic body, respectively.

In the present paper we consider matter described as a collisionless gas.
In astrophysics, this model is used to analyze
galaxies or globular clusters
where the stars play the role of the gas particles and collisions among
these are sufficiently rare to be neglected. The particles
only interact by the gravitational field which the ensemble creates
collectively, and the general relativistic description of such an ensemble
is given by the Einstein-Vlasov system.
The existence of spherically symmetric steady states to this system
has for example been shown in \cite{RR00}. In \cite{AKR} the present
authors proved the existence of static, axially symmetric solutions which
are no longer spherically symmetric, but which still have zero
angular momentum. In the present paper we also remove the latter
restriction, a task which, in view of what was said above, is
not trivial.

We shall formulate the Einstein-Vlasov system
in standard axial coordinates $t\in \R,\ \rho \in [0,\infty[,\
z\in \R,\ \varphi\in [0,2 \pi]$. Following \cite{Bard},
we write the metric in the form
\begin{equation} \label{metric_ax}
   ds^2=-c^2 e^{2\nu/c^2} dt^2 + e^{2\mu} d\rho^2 + e^{2\mu} dz^2+
   \rho^2 B^2 e^{-2\nu/c^2} (d\varphi-\omega dt)^2
\end{equation}
for functions $\nu, B, \mu, \omega$ depending on $\rho$ and $z$.
The reason for keeping the speed of light $c$ as a parameter
in the metric will become clear shortly.
The metric is to be asymptotically flat in the sense
that the boundary values
\begin{equation} \label{bc_infinity}
\lim_{|(\rho,z)| \to \infty} (|\nu|+|\mu| + |\omega|+ |B-1|)(\rho,z) = 0
\end{equation}
are attained at spatial infinity with certain rates which are specified later.
In addition we require that the metric is locally flat
at the axis of symmetry:
\begin{equation} \label{bc_axis}
   \nu(0,z)/c^2 + \mu(0,z) = \ln B(0,z),\ z\in \R .
\end{equation}
The solutions obtained in \cite{AKR} are static
and have zero total angular momentum.
In terms of the metric above this means that $\omega = 0$.
The quantity $\omega$ is the angular velocity with respect to infinity
of the invariantly defined zero angular momentum observers with worldlines
perpendicular to the $\{t={\rm const.}\}$ hypersurfaces; see \cite{Bard}.
For a rotating configuration, $\omega$ must not vanish identically.
As in \cite{AKR} the solutions to the Einstein-Vlasov system
are obtained by perturbing off a spherically symmetric Newtonian steady
state using two parameters, $\gamma = 1/c^2$ to turn on general relativity,
and $\lambda$ to turn on angular momentum.
In order to apply the implicit function theorem
and to make certain solution operators well-defined, it
becomes essential to handle the linearized $\omega$-equation
\begin{equation}\label{omeg-equ}
   \partial_{\rho\rho}\omega +\partial_{zz}\omega +\frac{3}{\rho}\,
   \partial_\rho \omega =q
\end{equation}
for a suitable class of right-hand sides $q$. The difficulty with this
fairly innocent looking elliptic equation is that the coefficient
$3/\rho$ blows up on the axis of symmetry, where
the full solution must remain smooth. It is technically
{\em very} demanding to make this equation as it stands fit into the
general framework of our approach, cf.\ the corresponding remark
in the appendix. However, this equation is
nothing but the Poisson equation on $\R^5$ where both $\omega$ and $q$ are
axially symmetric, i.e., they depend on $\rho=|(x_1,x_2,x_3,x_4)|$
and $z=x_5$. This observation turns out to make the inclusion of
non-trivial total angular momentum quite neat. We are not aware of a physical
background for this fact, nor are we aware that this observation
has previously been exploited in the area of mathematical relativity.
It turns out that an analogous observation applies to the
linearized equation for
$B$ which can be turned into the Poisson equation on $\R^4$.
This simplifies the proof and improves the result also for
vanishing angular momentum, when compared with \cite{AKR}.
One should realize that the generalization  to non-trivial angular
momentum means that one moves to a geometrically
more complex spacetime. To appreciate the fact that the resulting complications
are of a genuinely relativistic, geometric nature one should notice that
in \cite{Rein00}, where an analogous strategy
was used to obtain axially symmetric steady states
in the Newtonian case, i.e., for the Vlasov-Poisson system,
one and the same proof gives
static solutions with zero total angular momentum and stationary ones which
rotate, depending only on which particular ansatz function is chosen.

Let us now give a formulation of the Einstein-Vlasov system.
In a kinetic model
like the Vlasov equation the particle ensemble is described by its
distribution function $f\geq 0 $ which is defined on the tangent bundle $TM$
of the spacetime manifold $M$. Let $g_{\alpha \beta}$ denote
the Lorentz metric on the spacetime
and let $p^\alpha$ denote the canonical momentum coordinates
which correspond to the chosen coordinates $x^\alpha$ on $M$.
The Einstein field equations
\begin{equation} \label{einst_gen}
G_{\alpha\beta} = 8 \pi c^{-4} T_{\alpha\beta}
\end{equation}
are then coupled to the Vlasov equation
\begin{equation} \label{vlasov_gen}
p^\alpha \partial_{x^\alpha} f -
\Gamma^\alpha_{\beta \gamma} p^\beta p^\gamma \partial_{p^\alpha} f = 0
\end{equation}
via the definition of the energy momentum tensor
\begin{equation} \label{emt_gen}
T_{\alpha \beta}
= c^{-1} |g|^{1/2} \, \int p_\alpha p_\beta f \,\frac{dp^0 dp^1 dp^2 dp^3}{m}.
\end{equation}
Here $\Gamma^\alpha_{\beta \gamma}$ are the Christoffel symbols
induced by the metric,
$|g|$ denotes the modulus of its determinant, and $m>0$
is the rest mass of the particle with phase space coordinates
$(x^\alpha,p^\beta)$.
The characteristic system of the Vlasov equation
(\ref{vlasov_gen}) are the geodesic equations written as a first
order system on $TM$. For physical reasons we must require that
all particles move forward in time, i.e., $p^\alpha$ is a timelike,
future pointing vector on the support of $f$. Moreover, we make the
standard assumption that all particles have the
same rest mass which we normalize to unity.
The distribution function $f$ is then supported
on the mass shell
\begin{equation}\label{massshell}
PM = \{ g_{\alpha \beta} p^\alpha p^\beta = -c^2m^2 = -c^2\
\mbox{and}\ p^\alpha\ \mbox{is future pointing}\}
\subset TM.
\end{equation}
It is now important to realize that due to the presence of $\omega$
in the metric, i.e., due to the fact that we want to allow for non-trivial
angular momentum of the spacetime, the mass shell condition can in general
{\em not} be used to express $p^0$ by the remaining variables on $TM$.
It turns out that this can be done if and only if the Killing vector
$\partial/\partial t$ which corresponds to the time translation symmetry
is timelike everywhere, i.e.,
\begin{equation} \label{noergosphere}
- g(\partial/\partial t,\partial/\partial t)
=c^2 e^{2 \nu/c^2} - \rho^2 B^2 \omega^2 e^{-2\nu/c^2} > 0.
\end{equation}
For the solutions which we construct we do a priori not know
whether this property holds or not. We can if we wish make sure
that it does hold so that there is no ergosphere. The question
whether among the solutions we construct there are solutions that do have
an ergosphere is open. The vector field
$\partial/\partial t + \omega \partial/\partial \varphi$
is always timelike and can therefore be used to fix the time orientation
of the spacetime. For the solutions we construct,
\begin{equation} \label{futurepointing}
- g(\partial/\partial t+ \omega \partial/\partial \varphi,p^\alpha)
= c^2 e^{2 \nu/c^2} p^0 > 0
\end{equation}
on the support of $f$ so that all particle trajectories have future pointing
tangent vectors as desired.
We refer to \cite{And05} for more background on the Einstein-Vlasov system
and state our main result.
\begin{theorem}
There exist stationary solutions of the Einstein-Vlasov system
(\ref{einst_gen}), (\ref{vlasov_gen}), (\ref{emt_gen}) with $c=1$
such that
the metric is of the form (\ref{metric_ax}) and satisfies the
boundary conditions (\ref{bc_infinity}), (\ref{bc_axis}),
and where the total angular momentum is non-zero.
\end{theorem}
For the proof of this result the following observation is important.
The symmetries of the metric imply that the quantities
\begin{eqnarray*}
E
&:=& - g(\partial/\partial t, p^\alpha)
=
c^2 e^{2\nu/c^2} p^0 + \rho^2 B^2 \omega e^{-2\nu/c^2} (p^3 - \omega p^0), \\
L
&:=&
g(\partial/\partial \varphi, p^\alpha) =
\rho^2 B^2 e^{-2\nu/c^2} (p^3 - \omega p^0),
\end{eqnarray*}
are constant along geodesics; notice that
\[
E=c^2 e^{2\nu/c^2} p^0 +\omega L.
\]
Here $E$ can be thought of as a local or particle energy and $L$
is the angular momentum of a particle with respect to the axis
of symmetry. The requirement that $p^\alpha$ be future pointing
implies that $E>0$ on the support of $f$, provided (\ref{noergosphere})
holds, i.e., provided there is no ergosphere.
Up to regularity issues a distribution function $f$
satisfies the Vlasov equation if and only if it is constant along geodesics.
Hence we make the ansatz
\[
f = \Phi(E,L)\delta(m-1),
\]
and the Vlasov equation (\ref{vlasov_gen}) holds. The $\delta$ distribution
on the right hand side is to restrict $f$ to particles with rest mass
equal to unity; notice that the rest mass $m$ is conserved along
geodesics as well.
If we insert this ansatz into the definition
(\ref{emt_gen}) of the energy momentum tensor the latter
becomes a functional $T_{\alpha \beta} =  T_{\alpha \beta} (\nu,B,\mu,\omega)$
of the yet unknown metric functions
$\nu,B,\mu,\omega$, and we are left with the problem of solving the field
Einstein equations (\ref{einst_gen}) with this right hand side.

We will obtain the solutions by perturbing
off spherically symmetric steady states of the Vlasov-Poisson system
via the implicit function theorem;
the latter system arises as the Newtonian limit of the Einstein-Vlasov
system. We will specify conditions on the ansatz function
$\Phi$ above such that a two parameter family of axially symmetric
solutions of the Einstein-Vlasov system passes through the corresponding
spherically symmetric, Newtonian steady state. The parameter
$\gamma = 1/c^2$ turns on general relativity and a second parameter $\lambda$
turns on the dependence on $L$ and hence axial symmetry;
$L$ is not invariant under arbitrary rotations about
the origin, so if $f$ depends on $L$ the solution is
not spherically symmetric. Moreover, suitable assumptions on the
ansatz function $\Phi$ will force $\omega$ to be non-trivial so that
the solution rotates about the axis $\rho=0$ and has non-zero
total angular momentum. One should be careful to notice here that
there are always particles with non-zero angular momentum $L$
provided $f$ is non-trivial and smooth on the mass shell, but in 
general this does not imply that the total angular momentum of 
the whole spacetime is non-trivial. The scaling symmetry of 
the Einstein-Vlasov system can then be used to obtain the desired solutions
for the physically correct value of $c$, and not only for large $c$. 

The idea of employing the implicit function theorem to
obtain new equilibrium configurations of self-gravitating matter
distributions from known ones can be traced back to
L.~Lichtenstein, who investigated the existence of
non-relativistic, axially symmetric,
stationary, self-gravitating fluid balls \cite{Li1,Li2}.
His arguments were put into a rigorous and modern framework
in \cite{HeiligN} and extended to the general relativistic
set-up in \cite{HeiligGR}. As mentioned above, rotating elastic bodies
were considered in \cite{ABS09}.
Our approach significantly differs from \cite{ABS09,HeiligGR}
not only in the matter model, but also in that we use the explicit form
of the metric stated in (\ref{metric_ax}), together with a reduced
version of the Einstein field equations.

The outline of the paper is as follows.
The detailed formulation of our main result and the set-up for
the application of the implicit function theorem are stated in the next section.
In Section~\ref{outline} we then give a detailed outline of its proof.
The proof consists of several steps
and some of them are more or less identical to the corresponding steps
in \cite{AKR} and need not be repeated. However, the logical
structure of the present proof will be given in full detail.
In Section~\ref{matter-sect} we collect some properties of the matter terms
which will be needed throughout. Section~\ref{Fwelldef} contains
information on certain Newton potentials which
is then used to show that the operator to which we apply the implicit
function theorem is well defined.
Section~\ref{eehold-sect} explains
how a solution of the reduced field equations leads to a solution
of all the field equations.
In an appendix we collect a few general results
on the regularity of axially symmetric functions,
and we comment on solving the equation (\ref{omeg-equ}) without resorting
to the device of moving it into a higher dimension.


\section{Basic set-up and the precise result} \label{setup}

\setcounter{equation}{0}

We introduce the parameter
$\gamma = 1/c^2 \in [0,\infty[$.
In order to handle the mass shell condition effectively it is useful
to introduce new momentum variables
\begin{equation} \label{v_var}
v^0= e^{\gamma\nu}p^0,\ v^1 = e^{\mu} p^1,\ v^2 = e^{\mu} p^2,\
v^3 = \rho\,B e^{-\gamma \nu} (p^3-\omega p^0).
\end{equation}
This turns the mass shell condition for general $m$ into
\[
-c^2 m^2 = -c^2 (v^0)^2 + (v^1)^2 + (v^2)^2 +(v^3)^2
\ \ \mbox{or}\ \ (v^0)^2 = m^2 + \gamma |v|^2
\]
where $v=(v^1,v^2,v^3)\in \R^3$ and $|v|$ is the Euclidean norm on $\R^3$.
We eliminate $v^0$ by choosing the positive root which makes sure that
(\ref{futurepointing}) holds, i.e., all particles move forward in time.
With $m=1$ we find that
\[
E = c^2 e^{\gamma \nu} \sqrt{1+\gamma |v|^2} + \omega L,\
L=\rho B e^{-\gamma \nu} v^3.
\]
In particular
\[
E > c^2 e^{\gamma \nu} |v^3|/c - |\rho \omega B e^{-\gamma \nu}| |v^3| \geq 0
\]
provided the no-ergosphere condition (\ref{noergosphere}) holds.
The formula (\ref{emt_gen}) for the energy-momentum tensor
turns into
\begin{equation}\label{emt_v}
T_{\alpha\beta} =
\int_{\R^3} p_\alpha p_\beta \Phi(E,L)\frac{d^3v}{\sqrt{1+\gamma |v|^2}},
\end{equation}
where $p_\alpha$ has to be expressed via (\ref{v_var}). Here we first
express the four dimensional integral in (\ref{emt_gen}) in terms
of $(v^0,\ldots,v^3)$, replace the integration variable $v^0$ by $m$,
and then use the fact that $f$ is $\delta$ distributed with respect
to $m$. In what follows we view $f$ as a function on the mass shell.

In order to turn on or off angular momentum we introduce a second
parameter $\lambda\in \R$,
and in order to obtain the correct Newtonian limit for $\gamma=0$
we adjust the ansatz for $f$ as follows:
\begin{equation} \label{fansatz}
f = \phi\left(E-1/\gamma\right) \psi(\lambda, L).
\end{equation}
The important point here is that
\begin{equation} \label{limitE}
E-1/\gamma = \frac{e^{\gamma \nu}\sqrt{1+\gamma |v|^2} -1}{\gamma} + \omega L
\to \frac{1}{2} |v|^2 + \nu + \omega L
\ \mbox{as}\ \gamma \to 0;
\end{equation}
see $(\phi 2)$ below.
For $\gamma=0$ this limit is to replace the argument
of $\phi$ in (\ref{fansatz}).
We specify the conditions on the functions $\phi$ and $\psi$.

\smallskip

\noindent
{\bf Conditions on $\phi$ and $\psi$.}
\begin{itemize}
\item[($\phi 1$)]
$\phi \in C^1(\R)$ and there exists $E_0>0$ such that
$\phi(\eta)=0$ for $\eta \geq E_0$ and $\phi(\eta) >0$ for $\eta < E_0$.
\item[($\phi 2$)]
The ansatz $f(x,v)=\phi\left(\frac{1}{2}|v|^2 + U(x)\right)$,
$x,v \in \R^3$,
leads to a compactly supported steady state
$f_N$ of the Vlasov-Poisson
system, i.e., there exists a solution $U=U_N \in C^2(\R^3)$
of the semilinear Poisson equation
\[
\Delta U = 4 \pi \rho_N =
4 \pi \int\phi\left(\frac{1}{2}|v|^2 + U\right)\, dv,\
U(0)=0,
\]
$U_N(x) = U_N(|x|)$ is spherically symmetric, and the support of
$\rho_N \in C^2_c(\R^3)$ is the closed ball
$\overline{B}_{R_N}(0)$ where $U_N(R_N)=E_0$ and  $U_N(r)<E_0$
for $0 \leq r < R_N$,
$U_N(r)>E_0$ for $r > R_N$.
\item[($\phi 3$)] We have
\[ 6+4\pi r^2 a_N(r)>0,\quad r\in [0, \infty[, \]
where
\[
a_N(r):=\int_{\R^3}\phi'\Big(\frac{1}{2}\,|v|^2+U_N(r)\Big)\,dv.
\]
\item[($\psi$)]
$\psi \in C_c^\infty(\R^2)$ is compactly supported, $\psi \geq 0$,
$\partial_L \psi (\lambda, 0)=0$ for $\lambda\in\R$,
and $\psi(0, L)=1$ on an open neighborhood of the set
\[
\{ L=L(x,v) \mid (x,v)\in \mathrm{supp}\, f_N\}.
\]
\end{itemize}
For the Newtonian steady state
\[
\lim_{|x|\to \infty} U_N(x) =U_N(\infty) > E_0.
\]
The normalization condition $U_N(0)=0$ instead of
$U_N(\infty) =0$ is unconventional from the physics point
of view, but it has technical advantages.
Examples for ansatz functions $\phi$ which satisfy
($\phi 1$) and ($\phi 2$) are found in \cite{BFH,RR00},
the most well-known ones being the polytropes
\[
\phi(E):=(E_0-E)^k_+
\]
for $1<k<7/2$; here $E_0>0$ and $(\cdot)_+$ denotes the positive part.
In \cite[Sect.~7]{AKR} it is shown that ($\phi 3$) holds
for a subclass of the polytropes.

We recall that the metric (\ref{metric_ax})
was written in terms of the axial coordinates
$\rho\in [0, \infty[$, $z\in\R$, $\varphi \in [0,2 \pi]$.
In what follows we shall also use the corresponding
Cartesian coordinates
\[
x= (\rho\cos\varphi,\rho\sin\varphi, z) \in \R^3,
\]
and by abuse of notation we write $\nu (\rho,z)=\nu(x)$ etc.
It should be noted that
tensor indices always refer to the spacetime coordinates $t,\rho,z,\varphi$.
In Section~\ref{regularity} we collect the relevant information
on the relation between regularity properties of axially symmetric functions
expressed in these different variables.
We can now give a detailed formulation of our main result.

\begin{theorem} \label{main}
There exist $\delta >0$ and a two parameter
family
\[
(\nu_{\gamma,\lambda},B_{\gamma,\lambda},\mu_{\gamma,\lambda},
\omega_{\gamma, \lambda})_{(\gamma, \lambda) \in
[0,\delta[\times ]-\delta,\delta [ } \subset C^2(\R^3)^4
\]
with the following properties:
\begin{itemize}
\item[(i)]
$(\nu_{0,0},B_{0,0},\mu_{0,0}, \omega_{0, 0}) = (U_N,1,0,0)$
where $U_N$ is the potential
of the Newtonian steady state specified in $(\phi 2)$.
\item[(ii)]
If for $\gamma > 0$ a distribution function is defined
by Eqn.~(\ref{fansatz}) and a Lorentz metric by (\ref{metric_ax})
with $c=1/\sqrt{\gamma}$ then we obtain a solution of the
Einstein-Vlasov system (\ref{einst_gen}), (\ref{vlasov_gen}), (\ref{emt_gen})
which satisfies the boundary condition (\ref{bc_axis}) and
is asymptotically flat.
For $\lambda \neq 0$ this solution is not spherically symmetric,
and for appropriate choices of $\psi$ its total angular momentum is non-zero.
\item[(iii)]
If for $\gamma=0$ a distribution function is defined
by Eqn.~(\ref{fansatz}), observing (\ref{limitE}), this yields
a steady state of the Vlasov-Poisson system with gravitational
potential $\nu_{0,\lambda}$ which is not spherically symmetric
for $\lambda \neq 0$.
\item[(iv)]
In all cases the matter distribution is compactly supported
both in phase space and in space.
\end{itemize}
\end{theorem}

\noindent
{\bf Remarks.}
\begin{itemize}
\item[(a)]
The smallness restriction to $\lambda$ implies that the solutions
obtained are close to being spherically symmetric, and that their total angular
momentum is small.
\item[(b)]
The smallness restriction to $\gamma=1/c^2$ is undesired because
$c$ is, in a given set of units, a definite number.
However, if $(f,\nu,B,\mu,\omega)$ is a solution for some choice
of $c\in]0,\infty[$ then the rescaling
\begin{eqnarray*}
\tilde f(\rho,z,p^1,p^2,p^3)
&=&
c^{-3}
f(c \rho, c z,c p^1,c p^2,p^3),\\
\tilde \nu(\rho,z)
&=&
c^{-2} \nu (c \rho, c z),\\
\tilde B(\rho,z)
&=&
B (c \rho, c z), \\
\tilde \mu(\rho,z)
&=&
\mu (c \rho, c z), \\
\tilde \omega(\rho,z)
&=&
\omega (c \rho, c z),
\end{eqnarray*}
yields a solution of the Einstein-Vlasov system with $c=1$;
notice that this rescaling turns (\ref{noergosphere}) into the
corresponding condition with $c=1$.
\item[(c)]
The metric functions $\nu$ and $\mu$ do not satisfy the boundary
conditions (\ref{bc_infinity}), but
\begin{equation} \label{bc_infinity_shift}
\lim_{|(\rho,z)| \to \infty} \nu(\rho,z) = \nu_\infty,\
\lim_{|(\rho,z)| \to \infty} \mu(\rho,z) = -\nu_\infty/c^2,
\end{equation}
see Proposition~\ref{moreinfo}.
If we now abuse notation and redefine $\nu=\nu-\nu_\infty$,
$\mu=\mu+\nu_\infty/c^2$ and $\omega=\omega e^{-\nu_{\infty}/c^2}$
then the original condition (\ref{bc_infinity}) is restored and
the metric (\ref{metric_ax}) takes the form
\begin{eqnarray*}
ds^2
&=& -c^2 e^{2\nu/c^2} c_1^2dt^2 \\
&&
{}+
c_2^2\left(e^{2\mu} d\rho^2 + e^{2\mu} dz^2+
\rho^2 B^2 e^{-2\nu/c^2} (d\varphi-\omega c_1 dt)^2 \right)
\end{eqnarray*}
with constants $c_1, c_2>0$ which simply amounts to
a choice of different units of time and space.
By general covariance of the Einstein-Vlasov system
(\ref{einst_gen}), (\ref{vlasov_gen}), (\ref{emt_gen})
the equations still hold.
\item[(d)]
In the course of the proof of the theorem additional regularity
properties and specific rates at which the boundary values
at infinity are approached will emerge.
\end{itemize}

We will transform the problem of finding
the desired solutions into the problem of finding zeros of
a suitably defined operator. The Newtonian steady state
specified in ($\phi 2$) will be a zero of this
operator for $\gamma=\lambda=0$, and the implicit function theorem
will yield our result. 

The Einstein field equations are overdetermined, and we need to
identify a suitable subset of (a combination of) these equations
which suffice to determine  $\nu, B, \mu, \omega$,
and which are such that at the end of the day all the field equations
hold once the reduced system is solved. To do so
we introduce the auxiliary metric function
\[
\xi = \gamma \nu + \mu.
\]
Let $\Delta$ and $\nabla$ denote the Cartesian Laplace and gradient operator
respectively. Taking suitable combinations of the field equations one finds
that
\begin{eqnarray}
&&
\Delta \nu + \frac{\nabla B}{B} \cdot \nabla \nu
-\frac{1}{2} \rho^2 B^2 e^{-4\gamma\nu}|\nabla \omega|^2
=
4\pi \biggl[\gamma^2 e^{(2\xi - 4\gamma \nu)}
\left(T_{00}+2\omega T_{03}\right)\nonumber\\
&&\qquad\qquad\qquad
+ \gamma(T_{11} + T_{22})
+ e^{2\xi} \left(\frac{\gamma}{\rho^2 B^2} +
\gamma^2 \omega^2e^{- 4\gamma \nu}\right)T_{33} \biggr],
\qquad \label{nu_eqn}
\end{eqnarray}
\begin{eqnarray}
&&
\Delta B + \frac{\nabla \rho}{\rho}\cdot \nabla B =
8 \pi \gamma^2 B \left( T_{11} + T_{22}\right), \label{B_eqn}\\
&&
\Delta \omega + \left(2 \frac{\nabla \rho}{\rho}
+ 3 \frac{\nabla B}{B}
- 4 \gamma \nabla\nu \right)\cdot \nabla\omega
=
\frac{16\pi \gamma^2e^{2\xi}}{\rho^2 B^2}\left(T_{03} + \omega T_{33}\right),
\qquad \label{o_eqn}\\
&&
\left(1+\rho \frac{\partial_\rho B}{B}\right)\partial_\rho \xi -
\rho \frac{\partial_z B}{B} \partial_z \xi \nonumber \\
&&
\qquad\qquad
= \frac{1}{2\rho B} \partial_\rho(\rho^2 \partial_\rho B) -
\frac{\rho}{2 B}\partial_{zz} B + \gamma^2 \rho\left((\partial_\rho \nu)^2 -
(\partial_z \nu)^2\right)\nonumber\\
&&
\qquad\qquad\qquad
-\gamma \rho^3 B^2 e^{-4\gamma\nu}
\left((\partial_\rho\omega)^2-(\partial_z \omega)^2\right),
\qquad \label{xi_eqna}\\
&&
\left(1+\rho \frac{\partial_\rho B}{B}\right)\partial_z \xi +
\rho\frac{\partial_z B}{B} \partial_\rho \xi
\nonumber\\
&&
\qquad\qquad
= \frac{\partial_\rho(\rho \partial_z B)}{B} +
2 \gamma^2 \rho\, \partial_\rho \nu \partial_z \nu
+\frac{1}{2} \gamma \rho^3 B^2 e^{-4\gamma\nu}
\partial_\rho\omega \partial_z \omega .\label{xi_eqnb}
\end{eqnarray}
We write
\[
B = 1 + b.
\]
By taking a suitable combination of
(\ref{xi_eqna}) and (\ref{xi_eqnb}) we obtain equations
which contain only $\partial_\rho \xi$ or $\partial_z \xi$
respectively, and we chose the former.
In the above equations
the terms $T_{\alpha \beta}$ are functions of the unknown quantities
$\nu, b, \omega, \xi=\gamma \nu + \mu$ for which we therefore
have obtained the following reduced system of equations;
throughout, $\Phi_{ij}=\Phi_{ij}(\nu,B,\xi,\omega,\rho;\gamma,\lambda)$:
\begin{eqnarray}
& &
\Delta \nu=4\pi \left[\Phi_{00}
+ \gamma \Phi_{11} + 2 \gamma \omega \Phi_{03}+
e^{2\xi}\left(\gamma \frac{1}{\rho^2 B^2} +
\gamma^2 \omega^2e^{-4\gamma\nu}\right)\Phi_{33}\right]
\nonumber \\
& &
\hspace{5em} -\,\frac{1}{B} \nabla b \cdot \nabla \nu
+ \frac{1}{2} \rho^2 B^2 e^{-4\gamma\nu}|\nabla \omega|^2,
\quad \label{rnu_eqn} \\
& &
\Delta  b + \frac{\nabla \rho}{\rho}\cdot\nabla b
=8\pi \gamma^2 B \Phi_{11},
\label{rb_eqn} \\
& &
\left((1+b+\rho \partial_\rho b)^2 + (\rho \partial_z b)^2\right)\,
\partial_\rho \xi
\nonumber \\
& &
\hspace{3em}{} =\rho \partial_z b \left(\partial_z b + \rho \partial_{z\rho}b
+2\gamma^2 \rho B \partial_\rho \nu \partial_z \nu
+\frac{1}{2}\gamma \rho^3 B^3 e^{-4\gamma\nu}
\partial_\rho\omega \partial_z \omega \right)\nonumber\\
& &
\hspace{4.5em} +(1+b+\rho \partial_\rho b)
\biggl(\frac{\rho}{2} (\partial_{\rho\rho}b +\frac{2}{\rho} b - \partial_{zz}b)
+\gamma^2\rho B
\left((\partial_\rho \nu)^2 - (\partial_z \nu)^2\right)\nonumber\\
& &
\hspace{10em} -\,\gamma \rho^3 B^3 e^{-4\gamma\nu}
\left((\partial_\rho\omega)^2- (\partial_z \omega)^2\right) \biggr),
\label{rxi_eqn}\\
& &
\Delta \omega + 2 \frac{\nabla \rho}{\rho}\cdot\nabla\omega
=\frac{16\pi}{\rho^2 B^2}e^{2\xi}
\left(\gamma \Phi_{03} + \gamma^2\omega \Phi_{33}\right)
- \left(3 \frac{\nabla b}{B}-
4 \gamma \nabla\nu \right)\cdot \nabla\omega.\nonumber\\
&&
\label{ro_eqn}
\end{eqnarray}
We supplement this system with the boundary condition (\ref{bc_axis}),
which in terms of the new unknowns reads
\begin{equation} \label{bcxi_axis}
\xi(0,z)=\ln\left(1+b(0,z)\right).
\end{equation}
It remains to determine precisely the dependence of the
functions $\Phi_{\alpha\beta}$ on the unknown quantities
$\nu, b, \xi,\omega$.
The corresponding computation
uses the new integration variables
\[
\eta = \frac{e^{\gamma \nu}\sqrt{1+\gamma |v|^2} -1}{\gamma},\
s = B e^{-\gamma \nu} v^3.
\]
We also introduce the notation
\begin{eqnarray*}
   m &=& m(\eta,B,\nu,\gamma)
   =B e^{-\gamma \nu}\sqrt{\frac{e^{-2\gamma \nu}(1+\gamma \eta)^2 -1}{\gamma}},\\
   l &=& l(s,B,\nu,\gamma)
   =\frac{1}{\gamma}
   \left(e^{\gamma\nu}\sqrt{1+\gamma \frac{e^{2\gamma\nu}s^2}{B^2}}-1\right),
\end{eqnarray*}
and we obtain
\begin{eqnarray}
&&
\Phi_{00}( \nu,B,\xi,\omega,\rho;\gamma,\lambda)
=
\gamma^2 e^{2\xi - 4\gamma \nu} T_{00}\label{Phi00}\\
&&
\qquad =
\frac{2\pi}{B} e^{2\xi - 4\gamma \nu}
\int_{(e^{\gamma \nu}-1)/\gamma}^\infty
\int_{-m}^m
(1+\gamma \eta+\gamma \rho s \omega)^2
\phi(\eta+\rho s \omega) \psi(\lambda, \rho s)
\, ds\, d\eta\nonumber\\
&&
\qquad =
\frac{2\pi}{B} e^{2\xi - 4\gamma \nu}
\int_{-\infty}^\infty
\int_{l+\rho\omega s}^\infty
(1+\gamma \eta)^2
\phi(\eta) \psi(\lambda, \rho s)
\, d\eta\, ds,\nonumber\\&&
\Phi_{11}( \nu,B,\xi,\omega,\rho;\gamma,\lambda)
=
T_{11}+T_{22} \label{Phi11} \\
&&
\qquad =
\frac{2\pi}{B^3}e^{2\xi}
\int_{(e^{\gamma \nu}-1)/\gamma}^\infty
\int_{-m}^m (m^2-s^2)\,
\phi(\eta+\rho s \omega) \psi(\lambda, \rho s)
ds\, d\eta\nonumber\\
&&
\qquad =
\frac{2\pi}{B^3}e^{2\xi}
\int_{-\infty}^\infty
\int_{l+\rho\omega s}^\infty (m^2(\eta-\rho\omega s,B,\nu,\gamma)-s^2)\,
\phi(\eta) \psi(\lambda, \rho s)
d\eta\,ds,\nonumber\\
&&
\Phi_{33}( \nu,B,\xi,\omega,\rho;\gamma,\lambda)
=
e^{2\xi} T_{33}\label{Phi33}\\
&&
\qquad =
\frac{2\pi\rho^2}{B}\,e^{2\xi}
\int_{(e^{\gamma \nu}-1)/\gamma}^\infty
\int_{-m}^m s^2
\phi(\eta+\rho s \omega) \psi(\lambda, \rho s)\, ds\, d\eta\nonumber\\
&&
\qquad =
\frac{2\pi\rho^2}{B}\,e^{2\xi}
\int_{-\infty}^\infty
\int_{l+\rho\omega s}^\infty s^2
\phi(\eta) \psi(\lambda, \rho s)\, d\eta\, ds,
\end{eqnarray}
\begin{eqnarray}
&&
\Phi_{03}( \nu,B,\xi,\omega,\rho;\gamma,\lambda)
=
\gamma e^{2\xi} T_{03}\label{Phi03}\\
&&
\qquad =
-\frac{2\pi\rho}{B}\,e^{2\xi}
\int_{(e^{\gamma \nu}-1)/\gamma}^\infty
\int_{-m}^m
s\,(1+\gamma \eta+\gamma \rho s \omega)\,
\phi(\eta+\rho s \omega) \psi(\lambda, \rho s)
\, ds\, d\eta\nonumber\\
&&
\qquad =
-\frac{2\pi\rho}{B}\,e^{2\xi}
\int_{-\infty}^\infty
\int_{l+\rho\omega s}^\infty
s\,(1+\gamma \eta)\,
\phi(\eta) \psi(\lambda, \rho s)
\, d\eta\, ds;\nonumber
\end{eqnarray}
we recall that $T_{11}=T_{22}$.
In the course of the proof we will
benefit from both of these two different representations
of the matter terms.

We now define the function spaces in which we will obtain
the solutions of the system (\ref{rnu_eqn})--(\ref{bcxi_axis}).
By abuse of notation we write axially symmetric
functions as functions of $x\in \R^3$ or alternatively of
$\rho\geq 0$, $z\in\R$.
We fix $0<\alpha < 1/2$, $0<\beta<1$ and consider the Banach spaces
\begin{eqnarray*}
{\cal X}_1
&:=&
\Big\{\nu \in C^{2,\alpha}(\R^3)\mid
\nu(x)=\nu(\rho,z)\ \mbox{and}\
{\|\nu\|}_{{\cal X}_1} <\infty\Big\},\\
{\cal X}_2
&:=&
\Big\{b\in C^{3, \alpha}(\R^3) \mid
b(x)=b(\rho,z)\ \mbox{and}\
{\|b\|}_{{\cal X}_2} <\infty\Big\},\\
{\cal X}_3
&:=&
\Big\{\xi\in C^{1, \alpha}(Z_R) \mid
\xi(x)=\xi(\rho,z)\ \mbox{and}\
{\|\xi\|}_{{\cal X}_3} <\infty\Big\},\\
{\cal X}_4
&:=&
\Big\{\omega \in C^{2, \alpha}(\R^3) \mid
\omega(x)=\omega(\rho,z)\ \mbox{and}\
{\|\omega\|}_{{\cal X}_4} <\infty\Big\},
\end{eqnarray*}
where
\[
Z_R:= \{ x\in \R^3 \mid \rho < R\}
\]
is the cylinder of radius $R>0$, the quantity $R$ being defined in (\ref{Rdef})
below, and the norms are defined by
\begin{eqnarray*}
{\|\nu\|}_{{\cal X}_1}
&:=&
{\|\nu\|}_{C^{2, \alpha}(\R^3)}
+ {\|(1+|x|)^{1+\beta} \nabla \nu\|}_\infty,\\
{\|b\|}_{{\cal X}_2}
&:=&
{\|b\|}_{C^{3, \alpha}(\R^3)}
+ {\|(1+|x|)^{3} \nabla b\|}_\infty , \\
{\|\xi\|}_{{\cal X}_3}
&:=&
{\|\xi\|}_{C^{1, \alpha}(Z_R)},\\
{\|\omega\|}_{{\cal X}_4}
&:=&
{\|\omega\|}_{C^{2, \alpha}(\R^3)}
+ {\|(1+|x|)^{4} \nabla \omega\|}_\infty,
\end{eqnarray*}
and
\[
{\cal X} := {\cal X}_1\times {\cal X}_2\times {\cal X}_3\times {\cal X}_4,
\quad
{\|(\nu, b, \xi,\omega)\|}_{\cal X}
:= {\|\nu\|}_{{\cal X}_1}+{\|b\|}_{{\cal X}_2}
+ {\|\xi\|}_{{\cal X}_3}+{\|\omega\|}_{{\cal X}_4}.
\]
Here
${\|\cdot\|}_\infty$ denotes the
$L^\infty$-norm, functions in $C^{k, \alpha}(\Omega)$
have by definition continuous derivatives up to order $k$
and all their highest order derivatives are H\"older continuous with exponent
$\alpha$, and
\[
{\|g\|}_{C^{k, \alpha}(\Omega)}
:=
\sum_{|\sigma|\leq k} {\|D^\sigma g\|}_\infty
+
\sum_{|\sigma| =  k}
\sup_{x, y\in \Omega,\,x\neq y} \frac{|D^\sigma g(x)-D^\sigma g(y)|}{|x-y|^\alpha},
\]
where $D^\sigma$ denotes the derivative
corresponding to a multi-index $\sigma \in \N_0^3$.
It will be straightforward to extend $\xi$ to $\R^3$ once
a solution is obtained in the above space.

The condition ($\phi 2$) on the Newtonian
steady state implies that there exists $R>R_N>0$ such that
\begin{equation}\label{Rdef}
U_N(r) > (E_0+U_N(\infty))/2\ \mbox{for all}\ r > R.
\end{equation}
If
\[
{\|\nu - U_N\|}_\infty < |E_0-U_N(\infty)|/4\
\mbox{and}\ 0\leq \gamma < \gamma_0,
\]
with $\gamma_0>0$ sufficiently small, depending on $E_0$ and $U_N$,
then
\[
\frac{e^{\gamma \nu(x)} -1}{\gamma} > E_0\ \mbox{for all}\ |x| > R.
\]
Since $L$ is bounded on the support of $\psi$
this implies that there exists some $\delta_0>0$ such that
for all $(\nu,b,\xi,\omega;\gamma, \lambda)\in {\cal U}$
the matter terms resulting from (\ref{Phi00})--(\ref{Phi33})
are compactly supported in $B_R(0)$, where
\begin{eqnarray} \nonumber
{\cal U} & := & \{(\nu,b,\xi,\omega;\gamma, \lambda) \in
{\cal X}\times [0, \delta_0[ \times ]-\delta_0, \delta_0[ \mid
\\
& &
\hspace{2cm} {\|(\nu,b,\xi,\omega)-(U_N,0,0,0)\|}_{\cal X}<\delta_0\}.
\label{Udef}
\end{eqnarray}
In addition we require that $\delta_0>0$ is sufficiently small
so that $B=1+b > 1/2$ for all elements
in ${\cal U}$ and the factor in front of $\partial_\rho \xi$ in (\ref{rxi_eqn})
is larger than $1/2$.

\smallskip

\noindent
{\bf Remark.} If we want to make sure that the
no-ergosphere condition
(\ref{noergosphere}) holds for the solutions we construct
we redefine
\begin{equation} \label{normredef}
{\|\omega\|}_{{\cal X}_4} :=
{\|\omega\|}_{C^{2, \alpha}(\R^3)}
+ {\|(1+|x|)^{3} \omega\|}_\infty + {\|(1+|x|)^{4} \nabla \omega\|}_\infty.
\end{equation}
If $\delta_0>0$ is sufficiently small then
$\nu$ is close to the given Newtonian potential
$U_N$, $B$ is close to $1$, and due to the redefined norm
$\rho\,\omega$ is bounded so that
(\ref{noergosphere}) holds if $\gamma$ is sufficiently small.

\smallskip

\noindent
Now we substitute an element
$(\nu,b,\xi,\omega;\gamma, \lambda) \in {\cal U}$
into the matter terms defined in (\ref{Phi00})--(\ref{Phi03}).
With the right hand sides obtained in this way the equations
(\ref{rnu_eqn})--(\ref{ro_eqn}) can then be solved,
observing the boundary condition
(\ref{bcxi_axis}). In order to do so we need to be a little careful
with the definition of axial symmetry, because we will rewrite
different equations as equations on $\R^n$ with suitable, different
dimensions $n\geq 3$.

We call a function $u:\R^n \to \R$ {\em axially symmetric}
if it is invariant under all rotations about the $x_n$-axis, or
equivalently,
if there exists a function $\tilde u:[0,\infty[ \times \R \to \R$
such that
\[
u(x) = \tilde u(\rho,z),\ \mbox{where}\
\rho=\sqrt{x_1^1+\ldots +x_{n-1}^2}\ \mbox{and}\ z=x_n\ \mbox{for}\ x\in \R^n.
\]
Of course we will identify $u$ and $\tilde u$. For what follows
it is important that we can view an axially symmetric function
$u$ as a function defined on {\em any} $\R^n$ with $n\geq 3$.
In particular, we remark that
\[
\Delta_n u =
\partial_{\rho\rho} u + \frac{n-2}{\rho}\partial_\rho u + \partial_{zz} u =
\Delta u + (n-3)\frac{\nabla \rho}{\rho}\cdot \nabla u,
\]
where the left hand side refers to Cartesian coordinates on $\R^n$
and the right hand side to  $\R^3$.
In view of this relation,
equation (\ref{ro_eqn}), when rewritten in terms of $\rho$ and $z$, takes
the form (\ref{omeg-equ}). The latter equation can be handled directly
in these variables, cf.\ Lemma~\ref{omeq-solu}, but it is much more
efficient to
observe that this equation is nothing but a semilinear Poisson
equation on $\R^5$ for an axially symmetric function.
Similarly, (\ref{rb_eqn}) is nothing but
a semilinear Poisson equation on $\R^4$.

Equation (\ref{rnu_eqn}) and the properly rewritten
equations (\ref{rb_eqn}) and (\ref{ro_eqn}) are solved
(in terms of the right hand sides which of course contain the unknowns)
by the corresponding
Newton potentials, and (\ref{rxi_eqn}) can simply be integrated
in $\rho$. We define the corresponding solution operators by
\begin{eqnarray}
&&
G_1(\nu, b, \xi,\omega; \gamma, \lambda)(x)
:=
-\int_{\R^3} \left(\frac{1}{|x-y|}-\frac{1}{|y|}\right)\,M_1(y)\,dy
\nonumber\\
&&
\qquad {}+ \frac{1}{4 \pi}\int_{\R^3}
\left[\frac{\nabla b\cdot\nabla \nu}{B}
-\frac{1}{2}\rho^2 B^2 e^{-4\gamma\nu}|\nabla\omega|^2\right](y)
\frac{dy}{|x-y|},\ x\in \R^3,
\label{G1def} \\
&&
G_2(\nu, b, \xi,\omega; \gamma, \lambda)(x)
:=
-\frac{1}{\pi} \int_{\R^4} M_2(y)\frac{dy}{|x-y|^2},\ x\in \R^4,
\label{G2def} \\
&&
G_3(\nu, b, \xi,\omega; \gamma, \lambda)(\rho,z)
:=
\ln\left(1+ b(0,z)\right)
+ \int_0^\rho g(s,z)\,ds,\ 0\leq \rho < R,\nonumber\\
&& \label{G3def}\\
&&
G_4(\nu, b, \xi,\omega; \gamma, \lambda)(x)
:=
\frac{1}{8\pi^2}\!\int_{\R^5}\!\left[ \frac{\nabla b}{B}\cdot\nabla \omega
+ 4 \gamma \nabla\nu \cdot \nabla\omega -M_3 \right]\!(y)\frac{dy}{|x-y|^3},
\nonumber\\
&&\hspace{280pt}\ x\in \R^5. \label{G4def}
\end{eqnarray}
Here we put
\begin{eqnarray}
M_1
&:=&
\Big(\Phi_{00}+\gamma\Phi_{11}+2\gamma\omega\Phi_{03} \nonumber \\
&&
\hspace{1em}
+\Big(\gamma\,\frac{1}{\rho^2 B^2}+\gamma^2\omega^2 e^{-4\gamma\nu}\Big)\Phi_{33}\Big)
(\nu, B, \xi,\omega,\rho; \gamma, \lambda), \label{M1:def} \\
M_2
&:=&
\gamma^2 B\,
\Phi_{11}(\nu, B, \xi,\omega,\rho; \gamma, \lambda),
\label{M2:def} \\[1ex]
M_3
&:=&
\frac{16 \pi}{\rho^2 B^2}(\gamma \Phi_{03}+\gamma^2\omega\Phi_{33})
(\nu, B, \xi,\omega,\rho; \gamma, \lambda), \label{M3:def}
\end{eqnarray}
and
\begin{eqnarray*}
g
&:=&
\left((1+b+\rho\partial_\rho b)^2 + (\rho \partial_z b)^2\right)^{-1} \\
&&
\times\Bigg[\rho \partial_z b\Big(\partial_z b+ \rho \partial_{z\rho}b +
2 \gamma^2 \rho B \partial_\rho \nu \partial_z \nu
+\frac{1}{2}\,\gamma\rho^3 B^3
e^{-4\gamma\nu}\partial_\rho\omega\partial_z\omega\Big) \\
&&
\quad +\,(1+b+\rho \partial_\rho b)
\bigg(\frac{\rho}{2} (\partial_{\rho\rho}b +\frac{2}{\rho}\partial_\rho b
- \partial_{zz} b)
+\gamma^2\rho B
\left((\partial_\rho \nu)^2 - (\partial_z \nu)^2\right) \\
&&
{}\hspace{11em}
-\gamma\rho^3 B^3 e^{-4\gamma\nu}((\partial_\rho\omega)^2-(\partial_z\omega)^2)
\bigg)\Bigg].\qquad\
\end{eqnarray*}
Finally we define the mapping to which we are going to apply the
implicit function theorem as
\[
{\cal F}: {\cal U} \to {\cal X},\
(\nu, b, \xi,\omega; \gamma, \lambda)\mapsto
(\nu, b, \xi,\omega)
-(G_1,G_2,G_3,G_4)(\nu, b, \xi,\omega; \gamma, \lambda).
\]
In the next section we obtain the solutions in Theorem~\ref{main}
as a two parameter family of zeros of this mapping.
It should be noted that functions resulting from the operators
$G_i$ will all be axially symmetric so that they can all be viewed
as functions on $\R^3$ even if they are at first defined on different
$\R^n$'s.

\section{Outline of the proof}
\label{outline}
We check in a number of steps that the mapping ${\cal F}$
satisfies the conditions for applying the implicit function theorem.
Some of these steps turn out to be identical, or almost identical,
to the corresponding steps in~\cite{AKR} in which cases the details
will be left out.

\smallskip

\noindent
{\em Step 1.}\\
We need to check
that the mapping ${\cal F}$ is well defined. Since in this step
the presence of $\omega$ and also the somewhat different set-up
for the space ${\cal X}$ causes some differences compared with the
analysis in \cite{AKR}, we deal with this issue in some detail
in Section~\ref{Fwelldef}.

\smallskip

\noindent
{\em Step 2.}\\
The next step is to see that
\[
{\cal F}(U_N, 0, 0, 0; 0, 0)=0.
\]
This is due to the fact that for $\gamma=\lambda=0$ the choice
$b=\xi=\omega=0$
trivially satisfies (\ref{rb_eqn}), (\ref{rxi_eqn}), (\ref{ro_eqn}),
while (\ref{rnu_eqn})
reduces to
\[
\Delta \nu =
4\pi \Phi_{00}(\nu,1,0,0;0,0)
\]
with
\[
\Phi_{00}(\nu,1,0,0;0,0) = 4\pi \int_{\nu}^\infty \phi(\eta)
\sqrt{2(\eta - \nu)}\,d\eta =
\int_{\R^3} \phi\left(\frac{1}{2}|v|^2 + \nu\right)\, dv;
\]
notice that $b=0$ implies that $B=1$.
By ($\phi 2$), $\nu = U_N$
is a solution of this equation, and the fact that $U_N\in {\cal X}_1$
is part of what was shown in the previous step.
Notice further that $\psi(0,L)=1$
on the support of $f_N$ so that this factor, which vanishes for large
$L$ and formally makes the ansatz depend
on $L$ also in the Newtonian case, does not affect
the Newtonian steady state at all.

\smallskip

\noindent
{\em Step 3.}\\
The mapping ${\cal F}$ is continuous, and continuously Fr\'{e}chet
differentiable with respect to $(\nu, b, \xi,\omega)$.
Since the new element $\omega$
does not affect the proof for the static case in any essential way we
refer to \cite[Sect.~5]{AKR} for the details of this step.

\smallskip

\noindent
{\em Step 4.}\\
We have to show that the Fr\'{e}chet derivative
\[
{\cal L}:= D{\cal F}(U_N, 0, 0, 0; 0, 0): {\cal X} \to {\cal X}
\]
is one-to-one and onto. Indeed,
\[
{\cal L}(\delta\nu,\delta b,\delta\xi,\delta\omega)
= \Bigl(\delta\nu-L_1(\delta\nu)-L_2(\delta b,\delta\xi),
\delta b,
\delta\xi -L_3(\delta b),\delta\omega\Bigr)
\]
where
\begin{eqnarray*}
L_1(\delta\nu)(x)
&:=&
-\int_{\R^3} \left(\frac{1}{|x-y|}-\frac{1}{|y|}\right)\,
a_N (y) \delta\nu (y)\,dy,\\
L_2(\delta b,\delta\xi)(x)
&:=&
\frac{1}{4 \pi}\int_{\R^3}
\nabla \delta b(y)\cdot\nabla U_N(y)\,\frac{dy}{|x-y|},\\
&&
{}-2 \int_{\R^3} \left(\frac{1}{|x-y|}-\frac{1}{|y|}\right)\,
\rho_N (y) \delta\xi(y)\,dy,\\
L_3(\delta b)(x)
&:=&
\delta b (0,z) +
\frac{1}{2}
\int_0^\rho s\,\left(\partial_{\rho\rho} \delta b +
\frac{2}{\rho}\partial_\rho \delta b-\partial_{zz} \delta b\right)(s,z)\, ds,\\
&&
\hspace{200pt} 0\leq \rho < R,
\end{eqnarray*}
with $a_N$ as defined in ($\phi 3$).
To see that ${\cal L}$ is one-to-one let
${\cal L}(\delta\nu,\delta b,\delta\xi,\delta\omega)=0$.
The second and the last component  of this identity imply that
$\delta b = 0 = \delta\omega$,
and hence also $\delta\xi=0$ by the third component.
It therefore remains to show that
$\delta\nu=0$ is the only solution of the equation
$\delta\nu = L_1(\delta\nu)$, i.e., of the equation
\[
\Delta \delta\nu = 4 \pi a_N \delta\nu,\ \delta\nu(0)=0,
\]
in the space ${\cal X}_1$. Under the assumption on $a_N$
stated in ($\phi 3$) this was established in \cite[Sect.~7]{AKR}.

To see that ${\cal L}$ is onto let $(g_1,g_2,g_3,g_4) \in {\cal X}$
be given. We need to verify that there exists
$(\delta\nu,\delta b,\delta\xi,\delta\omega) \in {\cal X}$
such that
${\cal L}(\delta\nu,\delta b,\delta\xi,\delta\omega)=(g_1,g_2,g_3,g_4)$.
The second and fourth components of this equation simply say that
$\delta b=g_2$ and that $\delta\omega=g_4.$
Now $\delta b \in {\cal X}_2$ implies that $L_3(\delta b) \in {\cal X}_3$
by Lemma \ref{axreg}(b). Hence we can set $\delta \xi = g_3 +L_3(\delta b)$
to satisfy the third component of the `onto' equation,
and it remains to show that
\[
\delta\nu - L_1(\delta\nu) = g_1 + L_2(\delta b,\delta\xi)
\]
has a solution $\delta\nu \in {\cal X}_1$.
Firstly, $L_2(\delta b,\delta\xi)\in {\cal X}_1$.
The assertion therefore follows from the fact that
$L_1 :{\cal X}_1 \to {\cal X}_1$ is compact. We refer to \cite[Lemma~6.2]{AKR}
for the proof of this property.
It is at this point that we use the fact that in
${\cal X}_1$ the decay assumption is weaker than what we actually get
for $G_1$ and that $0<\alpha<1/2$: $L_1$ gains some H\"older
regularity which is the source for the compactness.

In view of the steps above we
can now apply the implicit function theorem,
cf.\ \cite[Thm.~15.1]{deimling}, to the mapping
${\cal F}: {\cal U} \to {\cal X}$.
\begin{theorem}\label{reducthm}
There exists
$\delta_1, \delta_2 \in ]0, \delta[$ and a unique, continuous solution map
\[
S: [0, \delta_1[\times ]-\delta_1, \delta_1[
\to B_{\delta_2}(U_N,0,0,0)\subset {\cal X}
\]
such that $S(0, 0)=(U_N, 0, 0, 0)$ and
\[
{\cal F}(S(\gamma, \lambda); \gamma, \lambda)=0\
\mbox{for all}\
(\gamma, \lambda) \in [0, \delta_1[ \times ]-\delta_1, \delta_1[.
\]
\end{theorem}
The definition of ${\cal F}$ implies that for any
$(\gamma,\lambda)$ the
functions $(\nu,b,\xi,\omega)=S(\gamma,\lambda)$ solve
the equations (\ref{rnu_eqn})--(\ref{ro_eqn}) where
the last equation deserves some explanation. By construction,
$\omega$ satisfies the equation $\Delta\omega = q$ on $\R^5$
where $q$ is an abbreviation for the right hand side of (\ref{ro_eqn}).
Both $\omega$ and $q$ are axially symmetric, cf.\
Lemma~\ref{newpotdec}.
Hence $\omega$ and $q$ can be viewed as functions of
$\rho$ and $z$, and as such they satisfy (\ref{omeg-equ})
which in turn implies that as functions on $\R^3$ they satisfy
(\ref{ro_eqn}). An analogous argument applies to (\ref{rb_eqn}).

If $f$ is defined by (\ref{fansatz}) then the equations
(\ref{nu_eqn})--(\ref{o_eqn}), (\ref{rxi_eqn}) hold with the induced
energy momentum tensor. We can extend $\xi$ to the whole space
using the solution operator $G_3$ for all $x\in \R^3$.
Also, the boundary condition
(\ref{bc_axis}) on the axis of symmetry is satisfied:
\[
\xi(0,z)= G_3(\nu,b,\xi)(0,z) = \ln (1+b(0,z))= \ln B(0,z);
\]
recall that $\xi = \gamma \nu + \mu$.
The solutions are asymptotically flat in view
of Remark (c) given after the formulation of Theorem~\ref{main};
also see Proposition~\ref{moreinfo}.

The solutions will in general have non-zero total angular momentum
for $\gamma\neq 0$. To prove this assume that $\omega = 0$.
Then
\[
T_{03} =
-\frac{2\pi\rho}{\gamma B}
\int_{(e^{\gamma \nu}-1)/\gamma}^\infty
(1+\gamma \eta)\,\phi(\eta)
\int_{-m}^m
s\,\psi(\lambda, \rho s) \, ds\, d\eta.
\]
It is easy to see that there are functions $\psi$ satisfying the
condition $(\psi)$ such that this integral is non-zero
in contradiction to (\ref{ro_eqn}) and $\omega=0$, e.g.,
choose $\psi$ such that on the support of $\psi$,
$\psi(\lambda,L)>\psi(\lambda,-L)$ for $L>0$ and $\lambda \neq 0$.

For $\gamma=0$ we conclude first that $b=0$, cf.~(\ref{rb_eqn})
and the $G_2$-part of the solution operator, respectively.
Then the $G_3$-part implies that $\xi=0$
and the $G_4$-part yields $\omega=0$,
so that the solution reduces to $(\nu,0,0,0)$, where $\nu$ solves
\[
\Delta \nu = 4 \pi \Phi_{00}(\nu,1,0,0,\rho;0,\lambda).
\]
Since
\[
\Phi_{00}(\nu,1,0,0,\rho;0,\lambda) = 4\pi
\int_{\nu}^\infty \phi(\eta+\rho s\omega)
\int_0^{\sqrt{2(\eta - \nu)}} \psi(\lambda, \rho s)\, ds\, d\eta
\]
coincides with the spatial density
induced by the ansatz (\ref{fansatz})
for the Newtonian case, cf.~\cite[Lemma~2.1]{Rein00},
part (iii) of Theorem~\ref{main} is established.
The resulting Newtonian steady state may or may not rotate,
depending on the properties of $\psi$;
see \cite[Remark (b), p.~324]{Rein00}.

The matter terms are compactly supported in view of of the discussion
following (\ref{Rdef}).

To complete the proof of Theorem~\ref{main}
we must show that all the field equations
are satisfied by the obtained metric (\ref{metric_ax}).
The argument relies on the Bianchi identity
$\nabla_\alpha G^{\alpha \beta}=0$ which holds for the
Einstein tensor induced by any (sufficiently regular) metric,
and on the identity $\nabla_\alpha T^{\alpha \beta}=0$
which is a direct consequence of the Vlasov equation~(\ref{vlasov_gen});
$\nabla_\alpha$ denotes the covariant derivative corresponding
to the metric (\ref{metric_ax}). Due to the inclusion of the
$\omega$-equation we cannot refer to the corresponding
argument in~\cite{AKR},
and we provide the details of the argument in Section~\ref{eehold-sect}.

We conclude this outline of the proof of our main result
by collecting some additional information on the solution which we
obtain in the course of the proof and which shows that
the solutions are asymptotically flat.
\begin{proposition}\label{moreinfo}
Let $(\nu,b,\xi,\omega)=S(\gamma,\lambda)$ be any of the solutions
obtained in Theorem~\ref{reducthm} and define
$\mu := \xi -\nu/c^2$ and $B:=1+b$. Then $\xi \in C^{2,\alpha}(\R^3)$,
the limit
$\nu_\infty:=\lim_{|x|\to\infty} \nu(x)$ exists, and
for all $\sigma\in\N_0^3$ with $|\sigma|\leq 2$
and $x\in \R^3$ the following estimates hold:
\begin{eqnarray*}
|D^\sigma (\nu (x)-\nu_\infty)|
&\leq&
C (1+|x|)^{-(1+|\sigma|)},\\
|D^\sigma (B-1) (x)|
&\leq&
C (1+|x|)^{-(2+|\sigma|)},\\
|D^\sigma \xi (x)|
&\leq&
C (1+|x|)^{-(2+|\sigma|)},\\
|D^\sigma \omega(x)|
&\leq&
C (1+|x|)^{-(3+|\sigma|)}.
\end{eqnarray*}
In particular, the spacetime equipped with the metric
(\ref{metric_ax}) is asymptotically flat in the sense
that (\ref{bc_infinity_shift}) and, after a trivial change of coordinates,
also (\ref{bc_infinity}) holds; see Remark (c) after Theorem~\ref{main}.
\end{proposition}
The proof follows easily from the decay estimates
for various Newton potentials
which we establish in Section~\ref{Fwelldef}, and from the fact that
$\xi$ now satisfies the equations (\ref{xi_eqna}) and (\ref{xi_eqnb});
see also
\cite[Prop.~2.3]{AKR}. In passing we remark that the asymptotic
behavior stated in Proposition~\ref{moreinfo} agrees with
what is given in \cite{Bard}.
The outline of the proof of our main result,
Theorem~\ref{main}, is now complete.


\section{Regularity of the matter terms}
\label{matter-sect}

\setcounter{equation}{0}

In this section we investigate the regularity properties
of the functions $\Phi_{00}$, $\Phi_{11}$, $\Phi_{33}$, and $\Phi_{03}$,
and of the induced matter terms $M_1$, $M_2$, and $M_3$
from (\ref{M1:def}), (\ref{M2:def}), and (\ref{M3:def}).
\begin{lemma}\label{PhiMreg}
Let $\phi$ and $\psi$ satisfy the conditions ($\phi 1$), ($\phi 2$), and
($\psi$).
\begin{itemize}
\item[(a)]
The functions $\Phi_{00},\Phi_{33}$, and $\Phi_{03}$
have derivatives
with respect to $\xi,\nu,\omega, \rho$, and $B\in ]1/2,3/2[$ up to order two,
and these are continuous in
$\nu,\xi,B,\omega,\rho,\gamma,\lambda$. For $\Phi_{11}$ the same is true with
derivatives up to order three.
\item[(b)]
For $(\nu,\xi,b,\omega,\gamma,\lambda)\in{\cal U}$ we have
$M_1,\,M_2, \in C_c^{1,\alpha}(\R^3)$,
$M_3 \in C^{0,\alpha}_c(\R^3)$, and $M_1,M_2,M_3$ are axially symmetric.
\end{itemize}
\end{lemma}
{\bf Proof.} (a) Differentiability
with respect to $\xi$ is obvious
to any order. Concerning differentiability with respect to
$\nu$, $\omega$, $B$, and $\rho$, the integrals in the formulae for $\Phi_{ij}$
expressed by means of $l$ gain one derivative. Since $\phi\in C^1(\R)$
and $\psi\in C_c^\infty(\R^2)$, the $\Phi_{ij}$ have the desired
regularity. For $\Phi_{11}$ we have to observe the following fact.
If we differentiate the integrand
in the second form of (\ref{Phi11}) with respect to one of the relevant
parameters we obtain an expression which
has the same structure as the $\Phi_{ij}$ in general.
If we differentiate the integration boundary
$l+\rho\omega s$ this gets substituted for $\eta$
in the integrand and the term $m^2 - s^2$ vanishes.

(b) By the choice of $R$ in (\ref{Rdef}) and $\delta_0$ in (\ref{Udef})
the matter terms which result by substituting an element
from ${\cal U}$ into the $\Phi_{ij}$
are compactly supported.
By the definitions of the spaces ${\cal X}_j$ the functions
which are substituted into $\Phi_{ij}$ are axially symmetric and
at least in $C^{1,\alpha}(\R^3)$.
The expression $\Phi_{33}$ contains the
factor $\rho^2$ so
that the term $\Phi_{33}/\rho^2$, which is present in both $M_1$ and $M_3$,
lies in $C^{1,\alpha}(\R^3)$.
Thus $M_1$ and $M_2$ belong to $C^{1,\alpha}_c (\R^3)$. In order to establish
the assertion for $M_3$ we only need to consider the expression
$N(\rho,z)/\rho$ where
\[
N(\rho,z) =
\int_{-\infty}^\infty
\int_{l+\rho\omega s}^\infty
s\,(1+\gamma \eta)\,
\phi(\eta) \psi(\lambda, \rho s)
\, d\eta\, ds.
\]
We now think of $\nu,B,\omega$ as functions of $\rho\in \R$ and
$z\in \R$ which are even in $\rho$ and lie in $C^2(\R^2)$.
Hence $N\in C_c^2(\R^2)$, and $N$ is odd with respect to $\rho\in\R$
as can be seen by the change of variables $s\mapsto -s$.
This easily implies that $N/\rho$ is in $C_c^1(\R^3)$, cf.\
\cite[Lemma~3.2]{AKR},
and hence $M_3 \in C_c^1(\R^3)\subset C^{0,\alpha}_c(\R^3)$ as claimed;
notice also Lemma~\ref{axreg}. \prfe

\noindent
{\bf Remark.}
The additional regularity of $\Phi_{11}$ is needed for the Fr\'{e}chet
differentiability of $G_2$; notice that $G_2$ maps into $C^{3,\alpha}$.


\section{${\cal F}$ is well defined}
\label{Fwelldef}

\setcounter{equation}{0}

In this section we show that
${\cal F}(\nu,b,\xi,\omega;\gamma, \lambda) \in {\cal X}$
for $(\nu,b,\xi,\omega;\gamma, \lambda) \in {\cal U}$.
For the most part this is an
assertion on certain Newton potentials.
The following lemma collects the necessary information.
\begin{lemma} \label{newpotdec}
Let $0<\alpha, \delta <1$, $n\in \N$ with $n\geq 3$,
and $g\in C^{0,\alpha}(\R^n)$ with
\[
|g(x)| \leq C (1+|x|)^{-n-\delta},\ x \in \R^n.
\]
Define
\[
U(x):= - \int_{\R^n} \frac{g(y)}{|x-y|^{n-2}}\,dy,\ x\in \R^n .
\]
Then $U\in C^{2,\alpha}(\R^n)$, and for any $\sigma \in \N^n_0$
with $|\sigma| \leq 2$,
\[
|D^\sigma U(x)| \leq C (1+|x|)^{2-n-|\sigma|},\ x \in \R^n.
\]
If $g$ is axially symmetric, then so is $U$.
\end{lemma}
{\bf Proof.}
Since $g \in L^1\cap L^\infty (\R^n)$ and H\"older continuous,
$U\in C^{2,\alpha}(\R^n)$ with
\[
\nabla U(x) =  (n-2) \int \frac{x-y}{|x-y|^{n}}\,g(y)\, dy
\]
and
\beas
\partial_{x_i}\partial_{x_j} U(x)
&=&
(n-2)\alpha_n \delta_{ij} g(x)\\
&&
{} + (n-2)  \int_{|x-y|\leq d}\partial_{x_j}\left(\frac{x_i-y_i}{|x-y|^n}\right)\,
\left(g(y) - g(x)\right)\, dy\\
&&
{} + (n-2) \int_{|x-y| > d}\partial_{x_j}\left(\frac{x_i-y_i}{|x-y|^n}\right)\,
g(y)\, dy
\eeas
for $i,j\in \{1,\ldots,n\}$, $x\in \R^n$, and $d>0$,
cf.~\cite{GT};
$\alpha_n$ denotes the volume of the unit ball in $\R^n$.
We consider the decay of the gradient first:
\beas
|\nabla U (x)|
&\leq&
(n-2) \int_{|x-y| < |x|/2} \frac{|g(y)|}{|x-y|^{n-1}} dy
+ (n-2) \int_{|x-y| \geq |x|/2} \ldots\\
&\leq&
C \int_{|x-y| < |x|/2} (1+|y|)^{-n-\delta}\frac{dy}{|x-y|^{n-1}}
+ C\|g\|_1|x|^{1-n}\\
&\leq&
C (1+|x|/2)^{-n-\delta}\int_{|x-y| < |x|/2} \frac{dy}{|x-y|^{n-1}}
+ C |x|^{1-n}\\
&\leq&
C (1+|x|)^{-n-\delta} |x|/2
+ C |x|^{1-n}
\leq C |x|^{1-n}.
\eeas
The decay for $U$ follows in the same way. For the second order derivatives
we observe that the first term in the formula above decays
as required by assumption on $g$.
We choose $d=(1+|x|)^{-n/\alpha}$.
Then
\beas
&&
\left|\int_{|x-y|\leq d}\partial_{x_j}\left(\frac{x_i-y_i}{|x-y|^n}\right)\,
\left(g(y) - g(x)\right)\, dy\right|\\
&&
\qquad \leq
C \|g\|_{C^{0,\alpha}(\R^3)} \int_{|x-y|\leq d}\frac{dy}{|x-y|^{n-\alpha}}
=
C \int_0^d r^{\alpha-1} dr= C d^\alpha \\
&&
\qquad  = C (1+|x|)^{-n} .
\eeas
In order to estimate the remaining term
we consider $x\in \R^n$
with $|x|$ sufficiently large: $|x|/2 \geq (1+|x|)^{-n/\alpha}$.
Then
\beas
&&
\left|\int_{|x-y| > d}\partial_{x_j}\left(\frac{x_i-y_i}{|x-y|^n}\right)\,
g(y)\, dy\right|\\
&&
\qquad \leq
C \int_{d<|x-y| < |x|/2} \frac{|g(y)|\, dy}{|x-y|^n}
+ C \int_{|x-y| \geq |x|/2} \ldots\\
&&
\qquad \leq
C \int_{d<|x-y| < |x|/2} (1+|y|)^{-n-\delta}\frac{dy}{|x-y|^n}
+ C \|g\|_1 |x|^{-n}\\
&&
\qquad \leq
C (1+|x|/2)^{-n-\delta}\int_{d<|x-y| < |x|/2} \frac{dy}{|x-y|^n}
+ C |x|^{-n}\\
&&
\qquad \leq
C (1+|x|/2)^{-n-\delta} \ln\left(\frac{|x|}{2}(1+|x|)^{n/\alpha}\right)
+ C |x|^{-n}
\leq C |x|^{-n}.
\eeas
Assume now that $g$ is axially symmetric, i.e., the
function is invariant under rotations about the $x_n$-axis.
Then the equation $\Delta U = n (n-2)\alpha_n g$, which is satisfied by $U$
on $\R^n$, is invariant under these rotations. Since $U$ vanishes
at infinity, it is the unique solution of this equation and hence
axially symmetric as well. \prfe
\begin{lemma}\label{G1welldef}
Let $(\nu,b,\xi,\omega;\gamma, \lambda) \in {\cal U}$ and let
$G_1=G_1(\nu,b,\xi,\omega;\gamma,\lambda)$
be as defined in (\ref{G1def}). Then $G_1 \in {\cal X}_1$
\end{lemma}
{\bf Proof.}
The source term $M_1$ of the first part of $G_1$ is in $C^{0,\alpha}_c (\R^3)$
by Lemma~\ref{PhiMreg}~(b).
The definitions of ${\cal X}_1$, ${\cal X}_2$, and ${\cal X}_4$ imply that
the source term
\[
\frac{1}{B} \nabla b\cdot\nabla \nu
-\frac{1}{2}\rho^2 B^2 e^{-4\gamma\nu}|\nabla\omega|^2
\]
of the second part of $G_1$
lies in $C^{0,\alpha} (\R^3)$ and decays like
$(1+|x|)^{-3-\beta}$.
Moreover, both source terms
are axially symmetric. Hence Lemma~\ref{newpotdec} with $n=3$
implies the assertion. \prfe

We notice that in the proof above we did not need to use the full
available regularity.
\begin{lemma}\label{G4welldef}
Let $(\nu,b,\xi,\omega;\gamma, \lambda) \in {\cal U}$ and let
$G_4=G_4(\nu,b,\xi,\omega;\gamma,\lambda)$
be as defined in (\ref{G4def}). Then $G_4 \in {\cal X}_4$.
\end{lemma}
{\bf Proof.}
We can view the source term
\[
q:= M_3 - 3 \frac{\nabla b}{B}\cdot\nabla\omega
+4\gamma\nabla\nu\cdot\nabla\omega
\]
as an axially symmetric function both on $\R^3$ and on $\R^5$.
By Lemma~\ref{PhiMreg}~(b) and the definitions
of ${\cal X}_1$, ${\cal X}_2$, and ${\cal X}_4$, $q\in C^{0,\alpha}(\R^3)$
and hence also $q\in C^{0,\alpha}(\R^5)$. Moreover, the compact support
of $M_3$ and the decay estimates for $\nabla\nu$, $\nabla b$, and
$\nabla\omega$ in the corresponding spaces imply that $q$ decays
like
\[
(1+|x|)^{-5-\beta} \sim (1+\rho + |z|)^{-5-\beta}
\]
where $x\in \R^3$ or in $\R^5$ has axial coordinates $\rho$
and $z$.
Lemma~\ref{newpotdec} with $n=5$
implies the assertion. The latter is true
with the original definition of the norm in ${\cal X}_4$
as well as with (\ref{normredef}).
\prfe

\begin{lemma} \label{G2welldef}
Let $(\nu,b,\xi,\omega;\gamma, \lambda) \in {\cal U}$ and let
$G_2=G_2(\nu,b,\xi,\omega;\gamma,\lambda)$
be as defined in (\ref{G2def}). Then $G_2\in {\cal X}_2$.
\end{lemma}
{\bf Proof.}
The source term $M_2$
lies in $C^{1,\alpha}_c(\R^3)$, and since it is axially symmetric
we can equally well view it as a function in  $C^{1,\alpha}_c(\R^4)$.
Thus Lemma~\ref{newpotdec} with $n=4$ implies the assertion;
notice that here we can throw one derivative onto the source term
so that $G_2$ ends up in $C^{3,\alpha}$.
\prfe

It remains to see that also $G_3\in {\cal X}_3$, but the corresponding
proof is identical to the one in \cite[Lemma~4.2~(c)]{AKR}.



\section{All field equations hold}
\label{eehold-sect}

\setcounter{equation}{0}

\begin{lemma}\label{bia}
Let $(\nu,b,\mu=\xi-\gamma\nu,\omega)$ be one of the solutions obtained in
Theorem~\ref{reducthm}. Then the metric (\ref{metric_ax})
together with $f$ defined by (\ref{fansatz}) solve the full
Einstein-Vlasov system
(\ref{einst_gen}), (\ref{vlasov_gen}), and (\ref{emt_gen}).
\end{lemma}
{\bf Proof.}
For a metric of the form (\ref{metric_ax}) the components
$00$, $11$, $22$, $33$, $03$, and $12$ of the field equations
are nontrivial. We have so far obtained a solution
$\nu,B,\xi,\omega$ of the reduced system
(\ref{rnu_eqn}), (\ref{ro_eqn}), (\ref{rb_eqn}),
(\ref{rxi_eqn}).
We define $E_{\alpha\beta}:= G_{\alpha\beta} - 8 \pi c^{-4} T_{\alpha\beta}$
so that the Einstein field equations become $E_{\alpha\beta}=0$.
By the reduced system,
\begin{equation} \label{00-03-33}
E_{00} + c^2 e^{-2(\mu-\nu/c^2)}(E_{11} + E_{22}) +
2 \omega E_{03} +\left(\frac{c^2}{\rho^2 B^2} e^{4\nu/c^2}+\omega^2\right) E_{33} =0,
\end{equation}
\begin{equation}\label{03-33}
E_{03} + \omega E_{33} = 0,
\end{equation}
\begin{equation}\label{11-22}
E_{11} + E_{22} =0,
\end{equation}
\begin{equation}\label{11-12-22}
\left(1 + \rho \frac{\partial_\rho B}{B}\right)(E_{11}-E_{22})+
\rho \frac{\partial_z B}{B} E_{12} =0.
\end{equation}
The Vlasov equation implies that
$\nabla_\alpha T^{\alpha\beta} = 0$,
and $\nabla_\alpha G^{\alpha\beta} = 0$ due to the contracted Bianchi
identity where $\nabla_\alpha$ denotes the covariant
derivative corresponding to the metric
(\ref{metric_ax}). Hence $\nabla_\alpha E^{\alpha\beta}=0$.
We want to use these relations
for $\beta = 1$ and $\beta=2$ together with (\ref{00-03-33})--(\ref{11-12-22})
to show that $E_{\alpha\beta}=0$.
To do so we first rewrite the equations
(\ref{00-03-33})--(\ref{11-12-22}) in terms of $E^{\alpha\beta}$.
Then (\ref{03-33}) and (\ref{11-22}) turn into
\begin{equation}\label{u-00-03}
\omega E^{00} - E^{03} = 0,
\end{equation}
\begin{equation}\label{u-11-22}
E^{11} + E^{22} =0.
\end{equation}
Using these to eliminate $E^{22}$ and $E^{03}$ the equations
(\ref{00-03-33}) and (\ref{11-12-22}) become
\begin{equation} \label{u-00-33}
\left(c^2 e^{4\nu/c^2} -\rho^2 B^2 \omega^2\right) E^{00}+ \rho^2 B^2 E^{33} =0,
\end{equation}
\begin{equation}\label{u-11-12}
2 \left(1 + \rho \frac{\partial_\rho B}{B}\right) E^{11} +
\rho \frac{\partial_z B}{B} E^{12} =0.
\end{equation}
The two Bianchi equations mentioned above can be written in the form
\begin{eqnarray}
&& \label{bia1}
\partial_\rho (\rho E^{11}) + \partial_z(\rho E^{12})
+\left(4\partial_\rho \mu + \frac{\partial_\rho B}{B}\right)(\rho E^{11})
+\left(4\partial_z \mu + \frac{\partial_z B}{B}\right)(\rho E^{12})\nonumber\\
&&
\qquad\qquad{}+\rho\left(\Gamma^1_{00} + 2 \omega \Gamma^1_{03}\right) E^{00} +
\rho \Gamma^1_{33} E^{33} = 0,\\
&& \label{bia2}
\partial_\rho (\rho E^{12}) - \partial_z(\rho E^{11})
+\left(4\partial_\rho \mu + \frac{\partial_\rho B}{B}\right)(\rho E^{12})
-\left(4\partial_z \mu + \frac{\partial_z B}{B}\right)(\rho E^{11})\nonumber\\
&&
\qquad\qquad{}+\rho\left(\Gamma^2_{00} + 2 \omega \Gamma^2_{03}\right) E^{00} +
\rho \Gamma^2_{33} E^{33} = 0,
\end{eqnarray}
where (\ref{u-00-03}) and (\ref{u-11-22}) were used to eliminate
$E^{03}$ and $E^{22}$.

At this point there is a small subtlety concerning regularity.
By definition of ${\cal X}_3$ we have $\xi\in C^{1,\alpha}$,
and since this function satisfies (\ref{rxi_eqn}) also
$\partial_\rho \xi\in C^{1,\alpha}$. Hence
$\partial_\rho (\rho E^{11})$ and $\partial_\rho (\rho E^{12})$
exist classically, but a priori
this need not be true for $\partial_z(\rho E^{11})$ and
$\partial_z(\rho E^{12})$. However,
approximating $\xi$ by smooth functions the corresponding
Bianchi identities again hold, and passing to the limit,
(\ref{bia1}) and (\ref{bia2}) hold in the sense of distributions.
With the possible exception of $\partial_z(\rho E^{11})$
and $\partial_z(\rho E^{12})$
all the terms in (\ref{bia1}) and (\ref{bia2})
are continuous so that these identities show that the latter terms
are classical, continuous derivatives as well.

We use (\ref{u-00-33}) and (\ref{u-11-12}) to eliminate
$E^{00}$ and $E^{11}$ from the two Bianchi identities (\ref{bia1})
and (\ref{bia2}). The resulting two equations contain only
$E^{12}$ and its first order derivatives and $E^{33}$.
Eliminating the latter finally yields the following first order
partial differential equation for $\rho E^{12}$:
\begin{eqnarray*}
&&
\left(\left(1 + \rho\frac{\partial_\rho B}{B}\right)^2 +
\frac{1}{2}\left(\rho\frac{\partial_z B}{B}\right)^2\right)
\partial_\rho(\rho E^{12})\\
&&
\qquad \qquad \qquad {}-
\frac{1}{2}\left(1 + \rho\frac{\partial_\rho B}{B}\right)
\left(\rho\frac{\partial_z B}{B}\right)
\partial_z(\rho E^{12})
+ c(\rho,z)\  (\rho E^{12}) = 0.
\end{eqnarray*}
Here $c=c(\rho, z)$ is a continuous function on $[0,\infty[\times \R$
the form of which is of no further interest, and the equation holds
for $\rho\geq 0$.

We recall that by definition (\ref{Udef}) of the set ${\cal U}$
the quantity $1 + \rho \partial_\rho B/B$ is bounded away from zero
so that any characteristic curve of the above equation must intersect
the axis $\rho=0$ where
$\rho E^{12}=e^{-2\mu}(\partial_zB/B -\gamma \partial_z \nu -\partial_z\mu)$
vanishes due to (\ref{bc_axis}).
Hence $E^{12}=0$ on $[0,\infty[\times \R$.
By (\ref{u-11-12}) also $E^{11}=0$, and by
(\ref{u-11-22}) the same is true for $E^{11}$. If we eliminate
$E^{00}$ from (\ref{bia1}) using (\ref{u-00-33}) we find an equation
of the form $(1 + \rho \partial_\rho B/B) E^{33} = 0$ so that
$E^{33} = 0$. By (\ref{u-00-33}) this implies that $E^{00}=0$
provided we choose the smallness parameter in the definition
(\ref{Udef}) of the set ${\cal U}$ such that the coefficient
of $E^{00}$ in (\ref{u-00-33}) does not vanish which is
possible due to the estimates in Proposition~\ref{moreinfo}.
By (\ref{u-00-03}) we finally find that $E^{03} = 0$,
and hence all the Einstein equations $E_{\alpha\beta}=0$ hold.\prfe




\section{Appendix: On the regularity of axially symmetric functions,
and a remark on the $\omega$ equation}
\label{regularity}

\setcounter{equation}{0}

We first collect a some remarks concerning the regularity of
axially symmetric functions on $\R^n$, $n\geq 3$.
\begin{lemma} \label{axreg}
Let $u:\R^n \to \R$ be axially symmetric and
$u(x)=\tilde u(\rho,z)$ where
$\tilde u:[0,\infty[ \times \R \to \R$.
Let $k\in \{1,2,3\}$ and $\alpha\in]0,1[$.
\begin{itemize}
\item[(a)]
$u\in C^k(\R^n)\ \Longleftrightarrow \ \tilde u \in C^k([0,\infty[ \times \R)$
and all derivatives of $\tilde u$ of order up to $k$ which are
of odd order in $\rho$ vanish for $\rho=0$.
\item[(b)] $u\in C^{0,\alpha}(\R^n) \ \Longleftrightarrow \
\tilde u\in C^{0,\alpha}([0,\infty[ \times\R)$.
\end{itemize}
\end{lemma}
The case $n=3$ is already stated in \cite[Lemma~3.1]{AKR},
and the proof does not depend on the space dimension.

We conclude this paper by pointing out that the $\omega$ equation
(\ref{omeg-equ}) can also be solved directly in the original
variables $\rho$ and $z$. However, as the authors
had to learn, basing the analysis on the
following result makes it {\em much} harder.
\begin{lemma}\label{omeq-solu}
A solution to the equation (\ref{omeg-equ})
is given by
\[
\omega(\rho, z) =
\int_0^\infty\int_{\R}
{\cal K}(\rho,z,\tilde{\rho},\tilde{z})\,q(\tilde{\rho}, \tilde{z})
\,d\tilde{z}\,d\tilde{\rho},
\]
where
\[
{\cal K}(\rho,z,\tilde{\rho},\tilde{z})
=-\frac{1}{2\pi}\left(\frac{\tilde\rho}{\rho}\right)^{3/2}
Q_{1/2}\left(\frac{\rho^2 + \tilde\rho^2 +(z-\tilde z)^2}{2\rho\tilde\rho}
\right),
\]
and $Q_{1/2}$ is a half-integer Legendre function of the second kind.
\end{lemma}

\noindent
{\bf Sketch of proof.}
As explained above, we can interpret $q$ as an axisymmetric
function on $\R^5$ and rewrite (\ref{omeg-equ}) as the Poisson
equation on $\R^5$. Hence the solution can be represented as
\[
\omega(x) = -\frac{1}{8\pi^2} \int_{\R^5} \frac{q(y)}{|x-y|^3} dy,
\ x\in \R^5.
\]
If we let $x=(0,0,0,\rho,z)$ and introduce polar coordinates
for the integration, two integrals can be carried out explicitly,
and the fact that
\[
Q_{1/2}(\chi) = \int_0^\pi \frac{\sin^2\eta}{(\chi - \cos \eta)^{3/2}} d\eta
\]
gives the formula in terms of ${\cal K}$. \prfe


\end{document}